\documentclass[letterpaper, 10 pt, conference]{ieeeconf}  

\usepackage{amsmath}

\IEEEoverridecommandlockouts                              

\usepackage{graphicx} 

\title{\LARGE \bf
Evaluating Pruning Methods in Gene Network Inference
}

\author{Michael M. Saint-Antoine$^{1}$ and Abhyudai Singh$^{2}$
\thanks{*This work was not supported by any organization}
\thanks{$^{1}$Center for Bioinformatics and Computational Biology, University of Delaware, Newark, Delaware 19716, USA
        {\tt\small mikest@udel.edu}}%
\thanks{$^{2}$Electrical and Computer Engineering, University of Delaware, Newark, Delaware 19716, USA
        }%
}

\begin{document}

\maketitle
\thispagestyle{empty}
\pagestyle{empty}

\begin{abstract}

One challenge in gene network inference is distinguishing between direct and indirect regulation. Some algorithms, including ARACNE and Phixer, approach this problem by using pruning methods to eliminate redundant edges in an attempt to explain the observed data with the simplest possible network structure. However, we hypothesize that there may be a cost in accuracy to simplifying the predicted networks in this way, especially due to the prevalence of redundant connections, such as feed forward loops, in gene networks. In this paper, we evaluate the pruning methods of ARACNE and Phixer, and score their accuracy using receiver operating characteristic curves and precision-recall curves. Our results suggest that while pruning can be useful in some situations, it may have a negative effect on overall accuracy that has not been previously studied. Researchers should be aware of both the advantages and disadvantages of pruning when inferring networks, in order to choose the best inference strategy for their experimental context.

\end{abstract}

\section{INTRODUCTION}

\subsection{Biology Concepts}

Much of the complex functionality of cells is due to gene interaction networks (GINs), in which genes can influence the expression of other genes, or of themselves [1,2]. If a gene codes for a protein that increases the expression of another gene, the first gene is said to ``activate'' the second gene. This can occur, for example, if the protein produced by the first gene is a transcription factor for the second gene. If a gene increases its own expression, that act is referred to as ``self-activation''. If a gene codes for a protein that decreases the expression of another gene, the first gene is said to ``inhibit'' the second gene. This can occur, for example, if a gene codes for a protein that blocks the promoter region of another gene, preventing RNA polymerase from binding to it. If a gene decreases its own expression, that act is referred to as ``self-inhibition'' [1,2]. Activation and inhibition between genes can give rise to complex GINs that underlie many cellular processes [1,2]. 

Understanding these GINs is important for understanding many diseases. Cancer, for instance, is caused by a loss of regulation of cellular growth, which is regulated at the genetic level by GINs [1]. Knowing the structure of the network can help us to find potential drug targets (such as upstream regulators of oncogenes) [3].

\subsection{Abstraction into Graph Theory}
For simplicity, it can be helpful to think of gene networks in an abstract graph theory framework. The network of all genes and their interactions can be viewed as a graph in which the genes are nodes, and the interactions between them are edges [1,2].

   \begin{figure}[thpb]

      \centering

      \includegraphics[scale=.45]{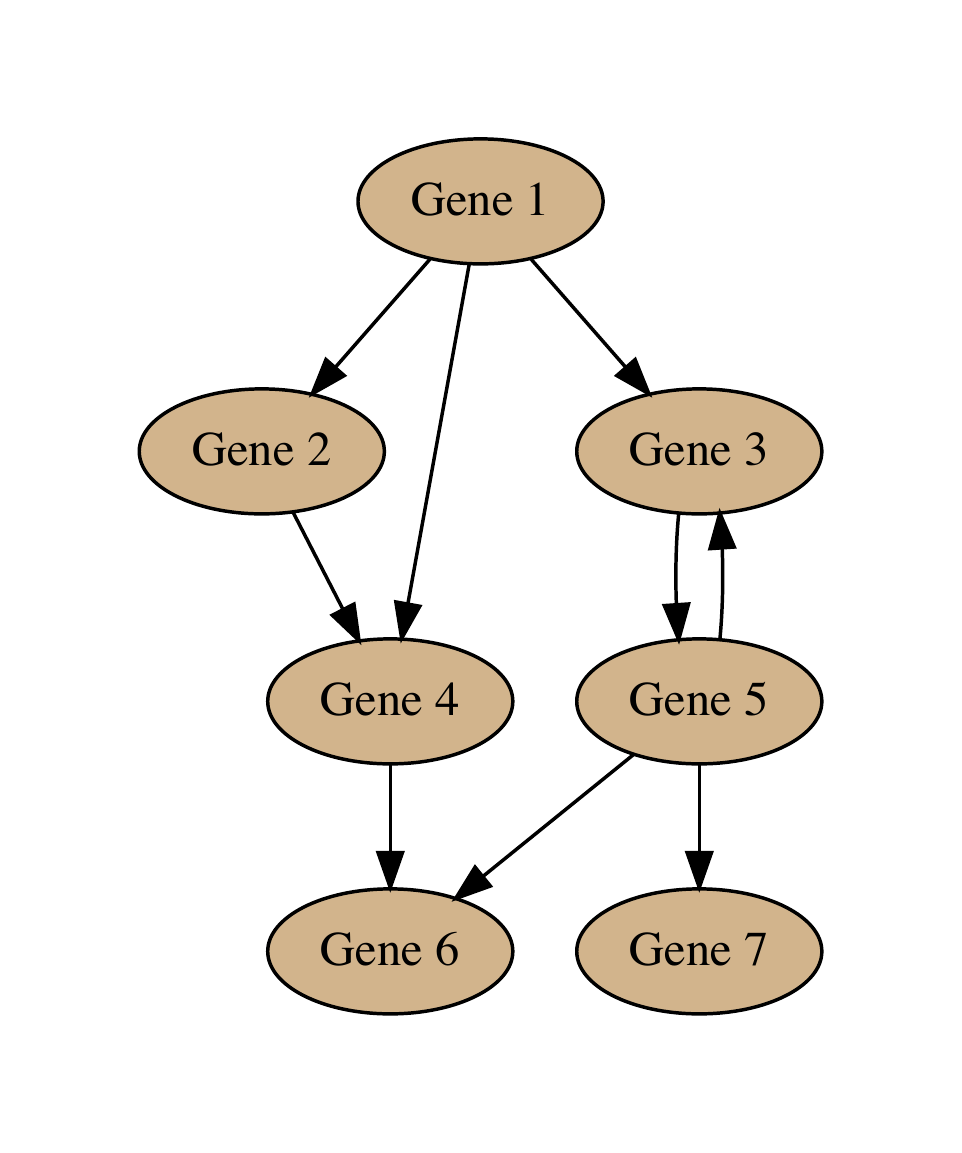}
      \caption{Example of an unsigned, directed network involving 7 genes.
}
      \label{figurelabel}
   \end{figure}

A graph is said to be undirected if for every edge X $\rightarrow$ Y there exists an equally weighted edge Y $\rightarrow$ X. In other words, undirected graphs make no attempt to predict the direction of causality in gene interactions. On the other hand, in a directed graph the edges X $\rightarrow$ Y and Y $\rightarrow$ X may have different weights, or one may exist independently, without the other. Both directed and undirected graphs are commonly used to describe GINs [1,2], but directed graphs are preferable (and more difficult to construct) as they describe the direction of causality in gene interactions, making them more biologically accurate.

A network can also be either signed or unsigned. A signed network distinguishes between activation edges and inhibition edges. An unsigned network makes no such distinction. In collaborations with biologists, it is important to produce a signed network to help them make sense of the interactions biologically. However, for theoretical research on network inference, it is common for unsigned networks to be used [4], since the difficult part of inferring a network structure is finding the edges, and once an edge is known, determining whether it is positive or negative becomes trivially easy (this can be found with a simple correlation or covariance measurement).

\subsection{Network Inference}
It can be difficult to experimentally detect interactions between genes. However, high-throughput sequencing techniques like RNA-Seq make it feasible to measure the expression levels of many different genes simultaneously [2]. Then, the challenge is to use computational techniques to analyze the expression data and find the gene interactions. This task is known as network inference.

There are several broad classes of network inference algorithms based on different computational techniques. Some of the most successful algorithms are based on information theory, conditional probability, and regression [2]. Simpler and computationally faster, though less accurate, techniques are based on correlation [2]. Bayesian techniques were popular at one point, but have fallen out of favor due to their inability to detect cycles [2,6]. ODE-based methods are also popular, but can only be used on time-series data [2].

\subsection{Evaluation}

To evaluate the performance of a network inference algorithm, it must be tested on expression data from a “gold standard” network, for which all of the edges are already known, and scored based on its ability to recover the true network structure from the input data. Fortunately for researchers, several gold standard networks have been made publicly available by Sage Bionetworks, a nonprofit organization that hosts network inference competitions called DREAM Challenges. In this paper, the DREAM5 Challenge Networks 1, 3, and 4 will be used for benchmarking [4]. Network 2 is omitted because it was not included in the publicly available DREAM5 download package.

Network 1 is an \textit{in silico} dataset containing expression data for 1643 genes. Network 3 is an \textit{E. coli} dataset containing expression data for 4511 genes. Network 4 is a \textit{S. cerevisiae} dataset containing expression data for 5950 genes. Each dataset also includes a list of ``potential transcription factors''. Some inference algorithms utilize these transcription factor lists while others do not. 

The output of an inference algorithm is a set of predicted edges, with each edge assigned a weight corresponding to its confidence level. From there, the researcher can choose which weight threshold to use for the final predicted network. It is typically the case that if one wants to increase the detection rate for true edges, then one must be willing to accept a higher false positive rate. Conversely, changing the threshold to decrease false positives will typically also decrease the true positive detection rate. When it comes to choosing a threshold, there is not necessarily a “right” or “wrong” answer, as different experimental contexts can lead to different priorities for balancing the maximization of true positives with the minimization of false positives. 

Despite the subjective nature of picking a threshold for the final network prediction, there are two commonly used metrics to objectively measure the accuracy of the inference algorithm output: receiver operating characteristic (ROC) curves and precision-recall (PR) curves [2,7]. There is currently some debate over which metric is better for evaluating performance, and a strong case has been made that the PR metric is superior in the GIN context [2,7]. However, since the official DREAM5 Challenge scoring methodology uses both metrics, we will also use both in this paper.

ROC curves describe the recall (true positive rate) of the predicted network as a function of the false positive rate. Equations 1 and 2 show the definitions of recall and false positive rate. TP represents the number of true positives. FP represents the number of false positives. TN represents the number of true negatives. FN represents the number of false negatives. 

\begin{equation}
Recall = \frac{TP}{TP+FN} 
\end{equation}

\begin{equation}
False\ Positive\ Rate = \frac{FP}{FP+TN} 
\end{equation}

The area under the ROC curve (AUROC) is typically used as a numerical accuracy score. A perfect inference algorithm would have an AUROC score of 1. However, no inference algorithm currently reaches this level of accuracy. AUROC scores for reasonably accurate inference algorithms typically fall within the 0.5 - 0.8 range [4], depending on how difficult the network is.

   \begin{figure}[h!]

      \centering

      \includegraphics[scale=.45]{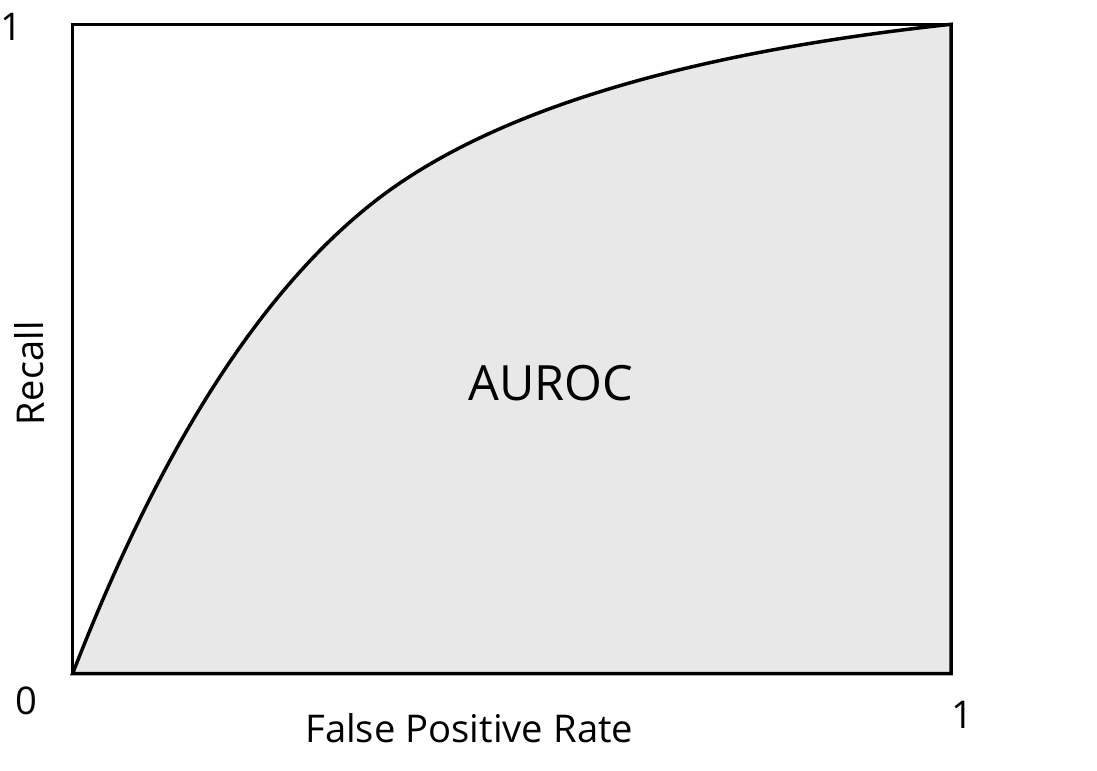}
      \caption{Receiver operating characteristic (ROC) curve example.
}
      \label{figurelabel}
   \end{figure}

PR curves describe the percentage of predicted edges that are correct (precision, defined in Equation 3 below) as a function of the percentage of true edges recovered (recall, defined in Equation 1). Area under the PR curve (AUPR) is used as a numerical accuracy score, and an AUPR of 1 is a perfect score.

\begin{equation}
Precision = \frac{TP}{TP+FP} 
\end{equation}

      \begin{figure}[h!]

      \centering

      \includegraphics[scale=.45]{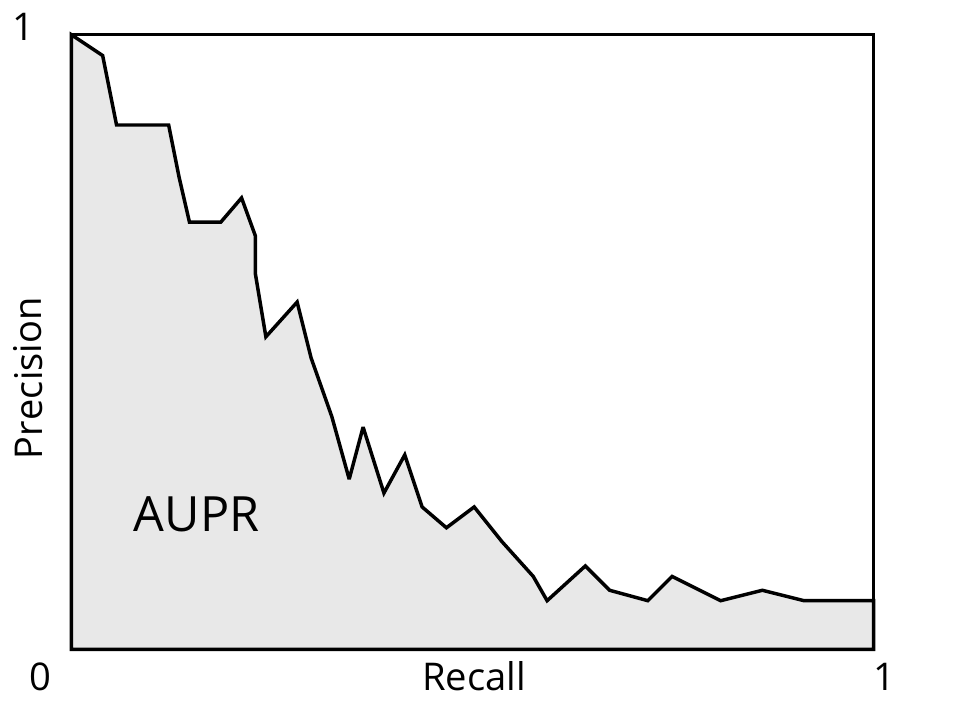}
      \caption{Precision-Recall (PR) curve example.
}
      \label{figurelabel}
   \end{figure}

All ROC and PR curvres in this paper were computed using the official DREAM5 Challenge scoring code, which has been made publicly available along with the gold standard networks [4]. For this scoring methodology, an edge prediction must have the correct direction to receive credit. So for a true edge X $\rightarrow$ Y, if an undirect algorithm predicts both edges X $\rightarrow$ Y and Y $\rightarrow$ X, then the former will be marked as a true positive while the latter will be marked as a false positive. However, even with this limitation, it is still possible for a highly accurate but undirected algorithm to perform well under this scoring protocol.

\section{Direct and Indirect Regulation}

\subsection{False Positive Problem}

One difficult challenge in network inference is distinguishing between direct and indirect regulation [5,6].

   \begin{figure}[h!]

      \centering

      \includegraphics[scale=.65]{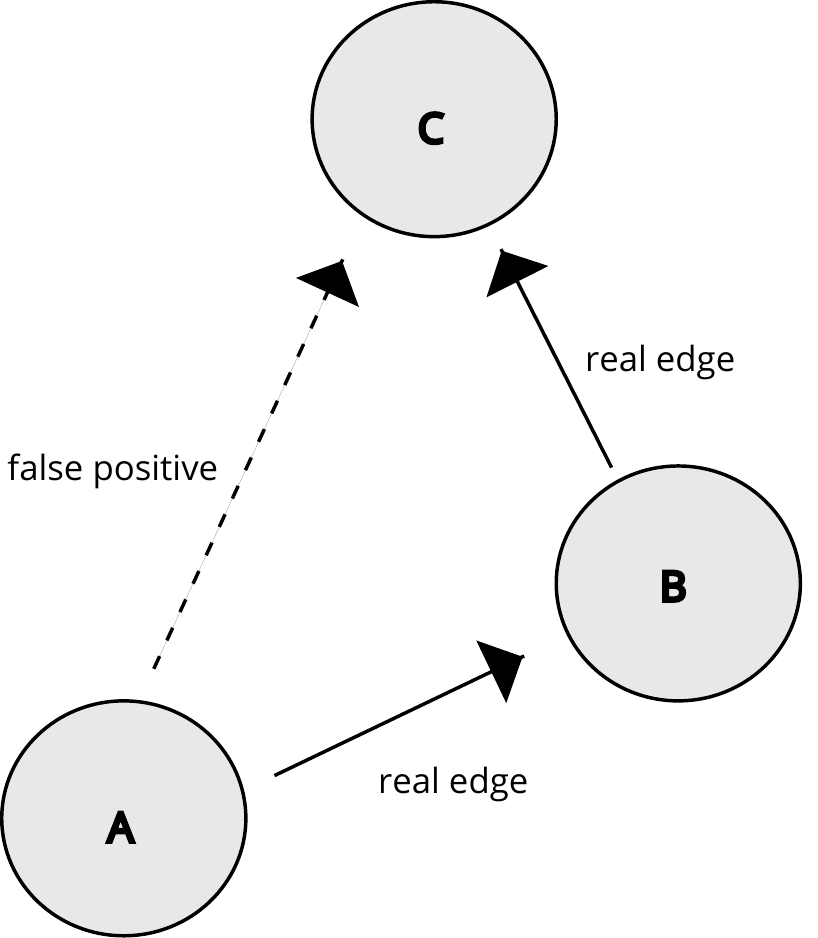}
      \caption{An indirect connection between A and C increases the chance of the edge A $\rightarrow$ C being falsely predicted.
}
      \label{figurelabel}
   \end{figure}

   In Figure 4, A regulates B, and B regulates C, but A does not directly regulate C. However, in many cases, the non-existent edge A $\rightarrow$ C will be more likely to be falsely predicted than two other randomly selected nodes, because of the indirect path between them. Some inference algorithms have attempted to correct for this problem by using pruning methods to eliminate redundant edges from the predicted network [5,6,9,10]. Much has been written about the potential advantages of these pruning methods, but their potential disadvantages have not yet been as widely studied.

\subsection{Feed Forward Loops}

One of the difficulties in correcting for the overprediction of edges between indirectly connected nodes is that feed-forward loops, in which a node both directly and indirectly regulates another node, are a common feature of many genetic networks [1]. The statistical overrepresentation of the feed forward loop network motif has been established in a variety of organisms [11,12]. The evolutionary reasons for this overrepresentation have also been studied, and it is thought that the functional role of feed forward loops is to allow for delayed response to a stimulus [13].

Feed forward loops present a challenge for the pruning methods discussed in the previous section, since they utilize the type of redundancy that the pruning methods are attempting to minimize. So, if an algorithm over-corrects the false positive problem, it could result in the under-prediction of real edges in feed-forward loops.

   \begin{figure}[h!]

      \centering

      \includegraphics[scale=.60]{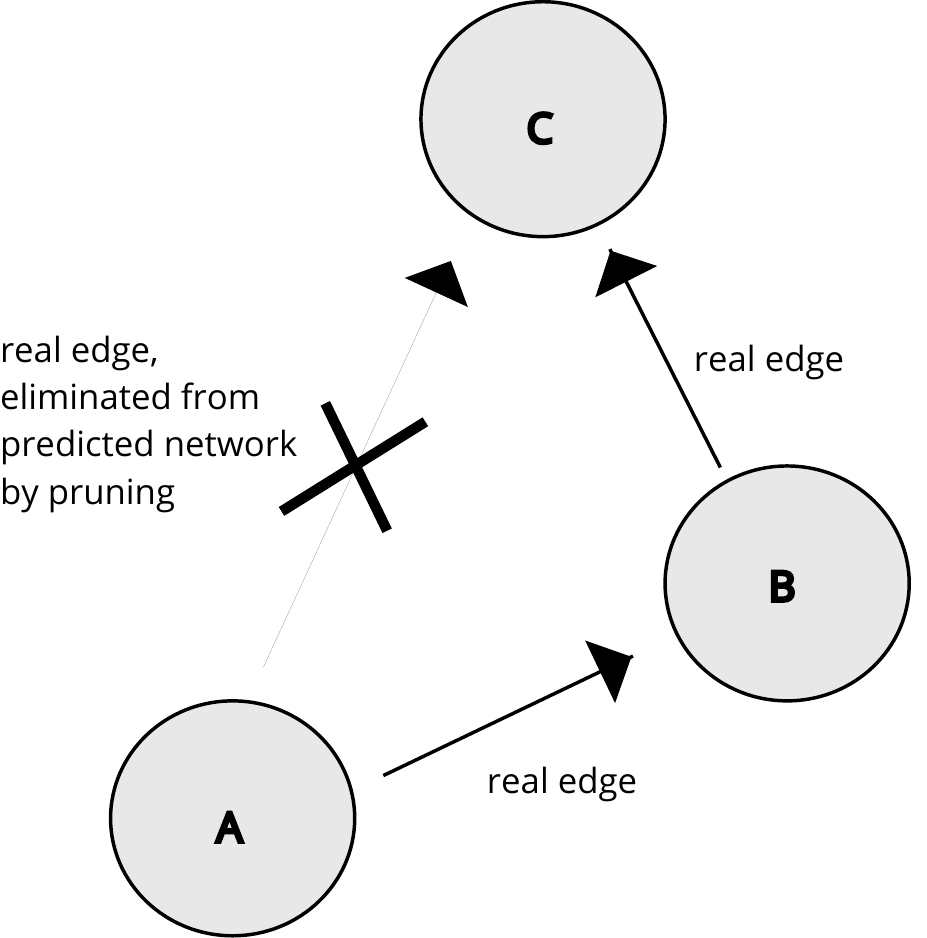}
      \caption{In a feed-forward loop with edges A $\rightarrow$ B, B $\rightarrow$ C, and A $\rightarrow$ C, a suboptimal pruning strategy could wrongly eliminate the edge A $\rightarrow$ C.
}
      \label{figurelabel}
   \end{figure}

Take, for example, Figure 5. Here, A regulates B, B regulates C, and A also regulates C. A pruning method based on minimizing redundancy could incorrectly identify the edge A $\rightarrow$ C as a false positive and eliminate it from the predicted network.

\section{Algorithms}

\subsection{ARACNE}

ARACNE [5], published in 2006, is one of the most tried and tested inference algorithms currently in use. Despite being more than a decade old, it is still commonly used and cited in many more recent papers in this field [2,6,8]. The algorithm is based on the information theory concepts of entropy and mutual information [5,6].

For a discrete random variable X, the entropy is defined as:

\begin{equation}
H(X)= - \sum_{x \in X}p(x)log(p(x)) 
\end{equation}

The logarithm here is usually base-2 in information theory, but a natural logarithm is used in ARACNE [5]. 

Entropy gives a numerical measure of the random variable's ``uncertainty'', and can also be thought of as the amount of new information gained when the value of a random variable becomes known.

Given two discrete random variables, X and Y, the mutual information between them is defined as:

\begin{equation}
I(X,Y)= H(X)+H(Y)-H(X,Y)
\end{equation}

\begin{equation}
= \sum_{x \in X}\sum_{y \in Y}p(x,y)log\frac{p(x,y)}{p(x)p(y)}
\end{equation}

If two random variables are independent, then their mutual information is 0. If two random variables are dependent, then the entropy of their joint distribution will be less than the sum of their individual entropies, giving them a positive mutual information level.

The ARACNE algorithm also contains several more complex computational steps on top of this information theory foundation. For a more in-depth explanation of the algorithm, please see the original paper: ``ARACNE: An Algorithm for the Reconstruction of Gene Regulatory Networks in a Mammalian Cellular Context'' (Margolin et al, 2006) [5].

A key feature of ARACNE is the use of the Data Processing Inequality (DPI) [5,6,8] to prune the network after the raw edge weights have been computed. For every possible triplet of nodes X, Y, and Z, the following inequality is checked:

\begin{equation}
I(X,Z)\leq \min \{ I(X,Y),I(Y,Z) \}
\end{equation}

If this statement is true, then the edge X $\rightarrow$ Z is eliminated. The goal of this pruning step is to yield a sparse network with minimal redundancy and maximal explanatory power.

\subsection{Phixer}

In the Phixer algorithm [6], developed by Nitin Singh in 2012, edge weights are computed using the phi-mixing coefficient. For random variables X and Y taking values in sets A and B, the phi-mixing coefficient $\phi(X|Y)$ is defined as:

\begin{equation}
\phi (X | Y)= \max_{S\subseteq A, T\subseteq B} | Pr \{ X \in S | Y \in T \}-Pr \{ X \in S  \} |
\end{equation}

$\phi(X|Y)$ is the weight of the edge Y $\rightarrow$ X. Since the phi-mixing coefficient is an asymetric measure, the weight of the edge X $\rightarrow$ Y may be different.

The Phixer algorithm includes its own pruning step, inspired by the DPI. For every possible triplet of nodes X, Y, and Z, the following inequality is checked:

\begin{equation}
\phi (X | Z) \leq \min \{ \phi (X | Y),\phi (Y | Z) \}
\end{equation}

If this statement is true, then the edge Z $\rightarrow$ X is eliminated. As with ARACNE, the goal is to yield a sparse network that can explain the observed data with the least redundancy.

\subsection{PIDC}

The Partial Information Decomposition and Context (PIDC) [8] algorithm was developed by Chan et al. in 2017, and is one of the more recent and better performing algorithms in the field of network inference. Although the PIDC paper references the DPI pruning method, the algorithm does not actually use the pruning method. However, we have included PIDC in order to compare the results from ARACNE and Phixer to a recently developed, highly accurate algorithm.

PIDC, like ARACNE, is based on the entropy and mutual information concepts seen in Equations 4, 5, and 6. However, instead of computing mutual information scores in a pairwise manner, PIDC computes the partial information decomposition between triplets of genes. For each target gene Z and source genes X and Y, the partial information decomposition is defined as:

\begin{equation}
\begin{split}
I(Z;X,Y)= Synergy(Z;X,Y)+Unique_{Y}(Z;X)  
\\+ Unique_{X}(Z;Y)+Redundancy(Z;X,Y)
\end{split}
\end{equation}

The ``Synergy'' term is the part of the information about Z that can be provided only by the combination of X and Y. The ``Unique'' terms are the parts of the information about Z that can be provided by only X or only Y. The ``Redundancy'' term is the part of the information about Z that can be provided by either X or Y, without needing to combine both of them.

This is only a very brief summary of the foundation of PIDC. For a more in-depth explanation of how to compute these terms and use them to calculate the final edge weights for the predicted network, please see the original paper ``Gene Regulatory Network Inference from Single-Cell Data Using Multivariate Information Measures'' (Chan et al, 2017) [8]. 

Please note that PIDC predicts an undirected network, so for every predicted edge X $\rightarrow$ Y, an equally weighted edge Y $\rightarrow$ X is also predicted. Despite this limitation, it still performs well when compared to directed algorithms.

\section{ANALYSIS}

\subsection{Overview}

We began our analysis by testing the ARACNE and Phixer algorithms on the DREAM5 Networks 1, 3, and 4 gold standard datasets. The algorithms were run both in their original form (with pruning), as well as with the pruning step removed, in order to compare the results. 

ARACNE was run without bootstrapping, using the suggested parameters from the official documentation [5], including taking the lists of potential transcription factors as an argument. After running it on each network in its original form, the pruning step was removed by adding the ``--nodpi'' option to the run command, and it was run on each network again without pruning.

Phixer was run with its default parameters, including 10 bootstrapping runs. It does not utilize the lists of transcription factors (although still performs reasonably well despite this). After running it on each network in its original form, we edited the code to remove the pruning step and then ran it on each network again.

PIDC was also run on each dataset, with its default parameters, for the purpose of comparing the ARACNE and Phixer results to those of a relatively new algorithm known for its high level of accuracy.

\subsection{Results}

Figures 6, 7, and 8 show the ROC curves for each network. The dotted lines are the curves for the unpruned versions of ARACNE and Phixer, while the solid lines are the curves for the pruned versions of these algorithms, as well as for PIDC.

   \begin{figure}[thpb]

      \centering

      \includegraphics[scale=.23]{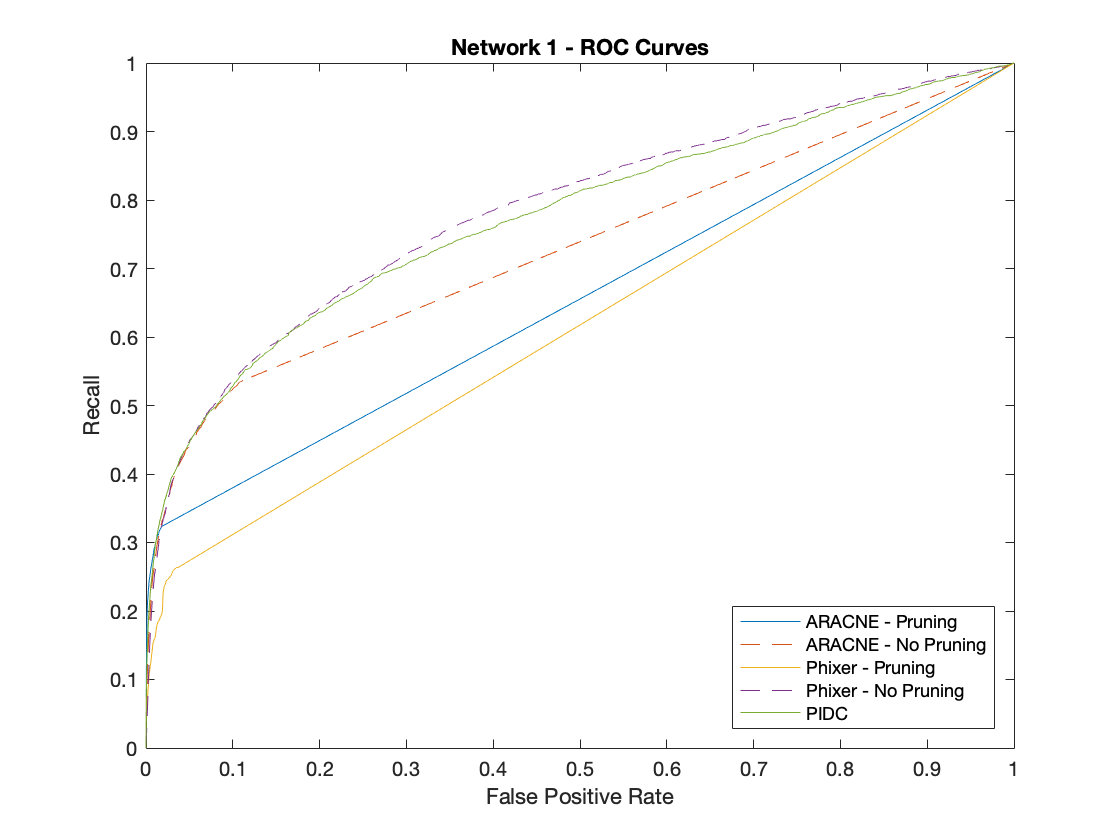}
      \caption{Network 1 ROC curves.
}
      \label{figurelabel}
   \end{figure}

      \begin{figure}[thpb]

      \centering

      \includegraphics[scale=.23]{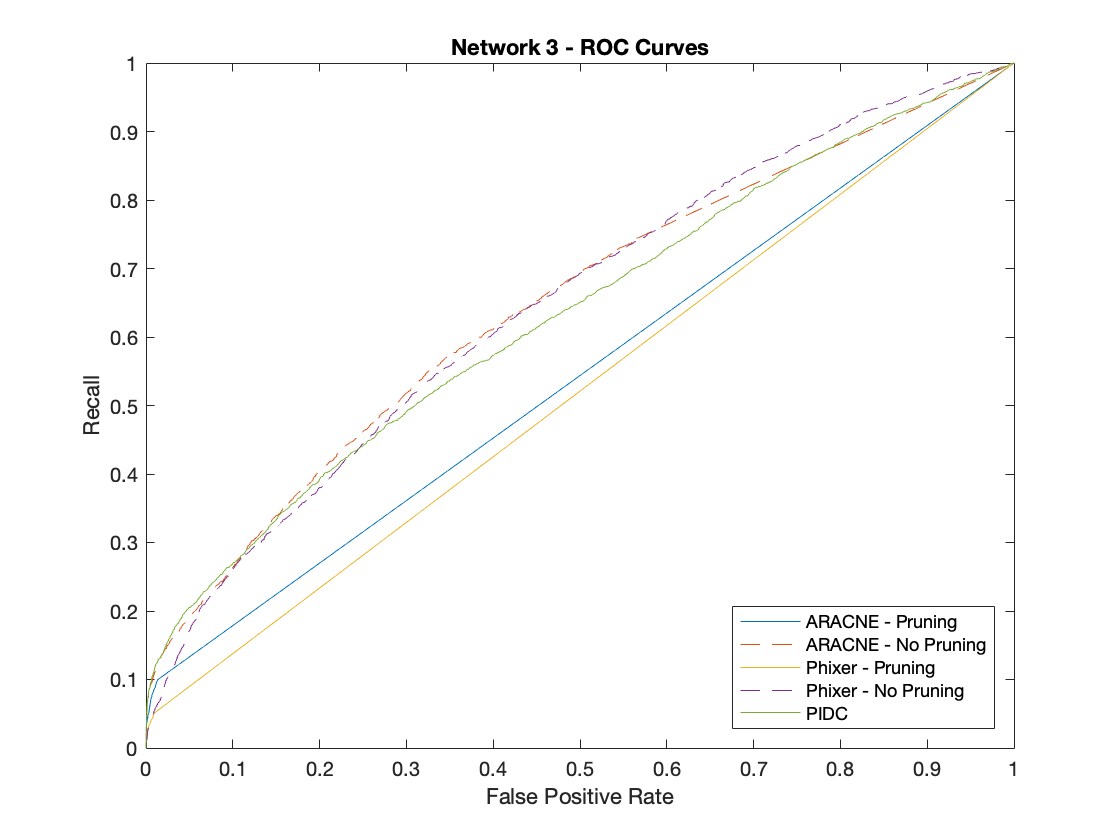}
      \caption{Network 3 ROC curves.
}
      \label{figurelabel}
   \end{figure}

         \begin{figure}[thpb]

      \centering

      \includegraphics[scale=.23]{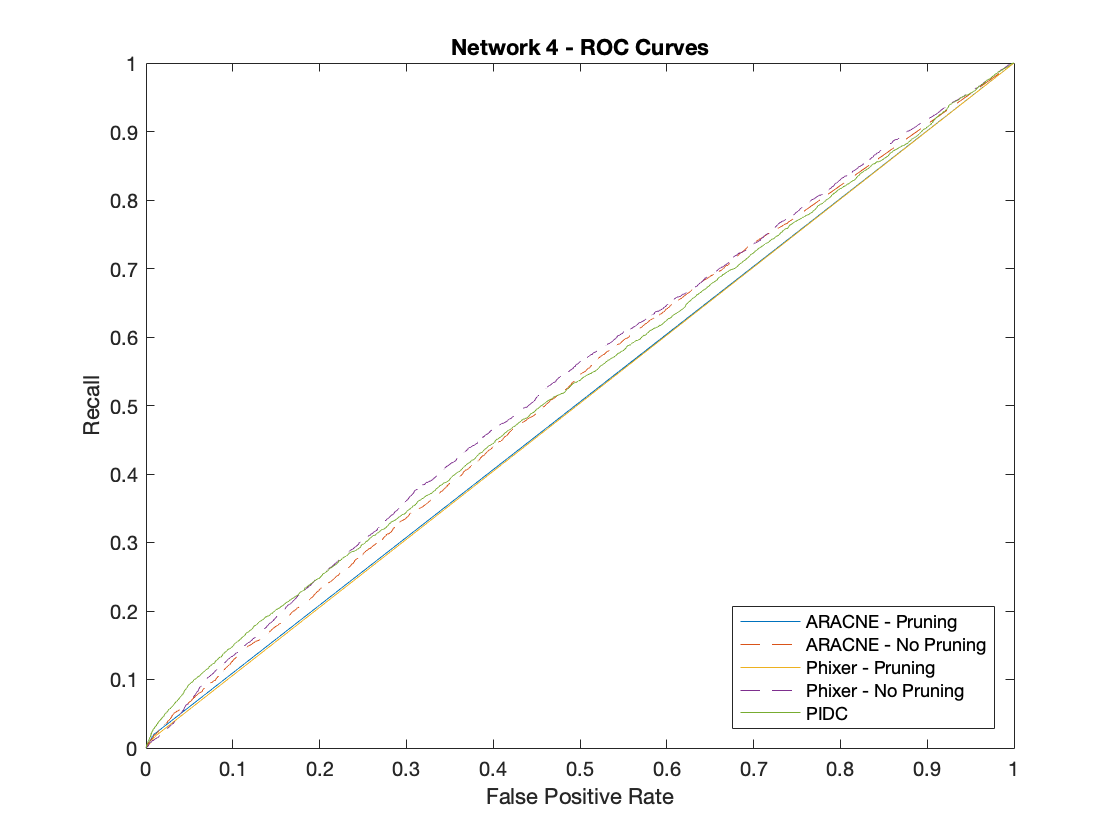}
      \caption{Network 4 ROC curves.
}
      \label{figurelabel}
   \end{figure}

         \begin{figure}[thpb]

      \centering

      \includegraphics[scale=.23]{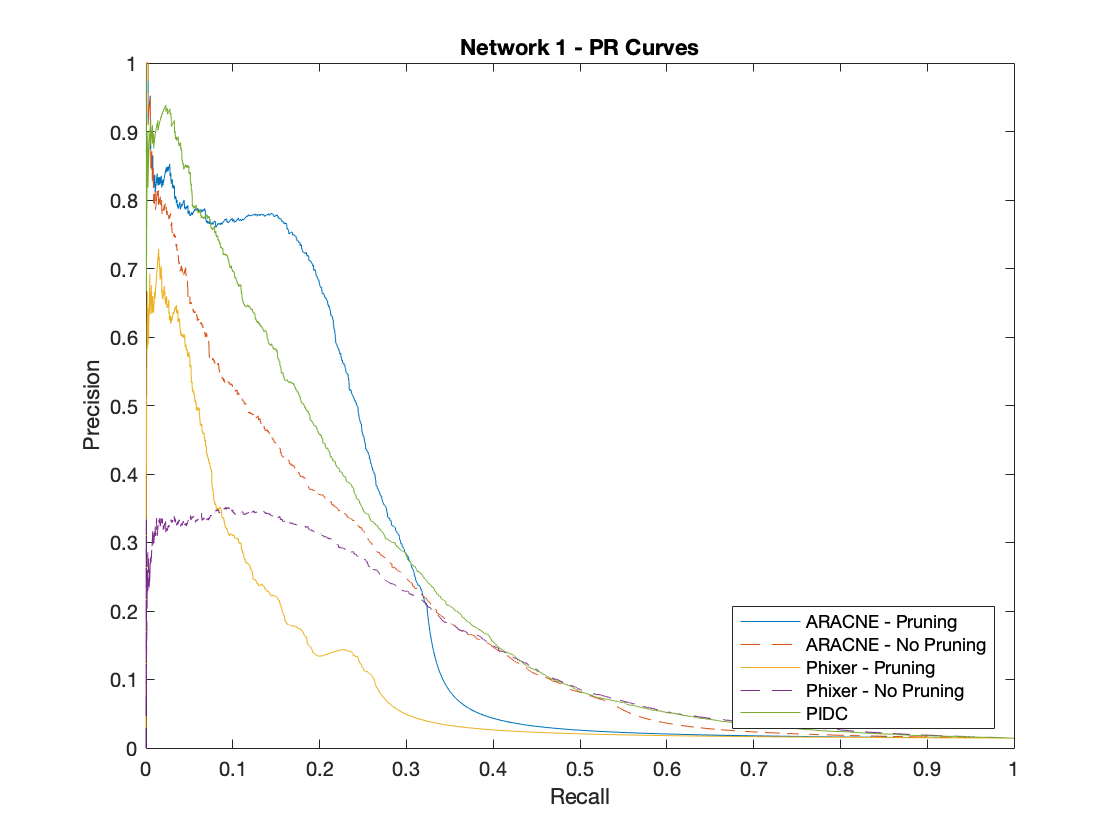}
      \caption{Network 1 PR curves.
}
      \label{figurelabel}
   \end{figure}

            \begin{figure}[thpb]

      \centering

      \includegraphics[scale=.23]{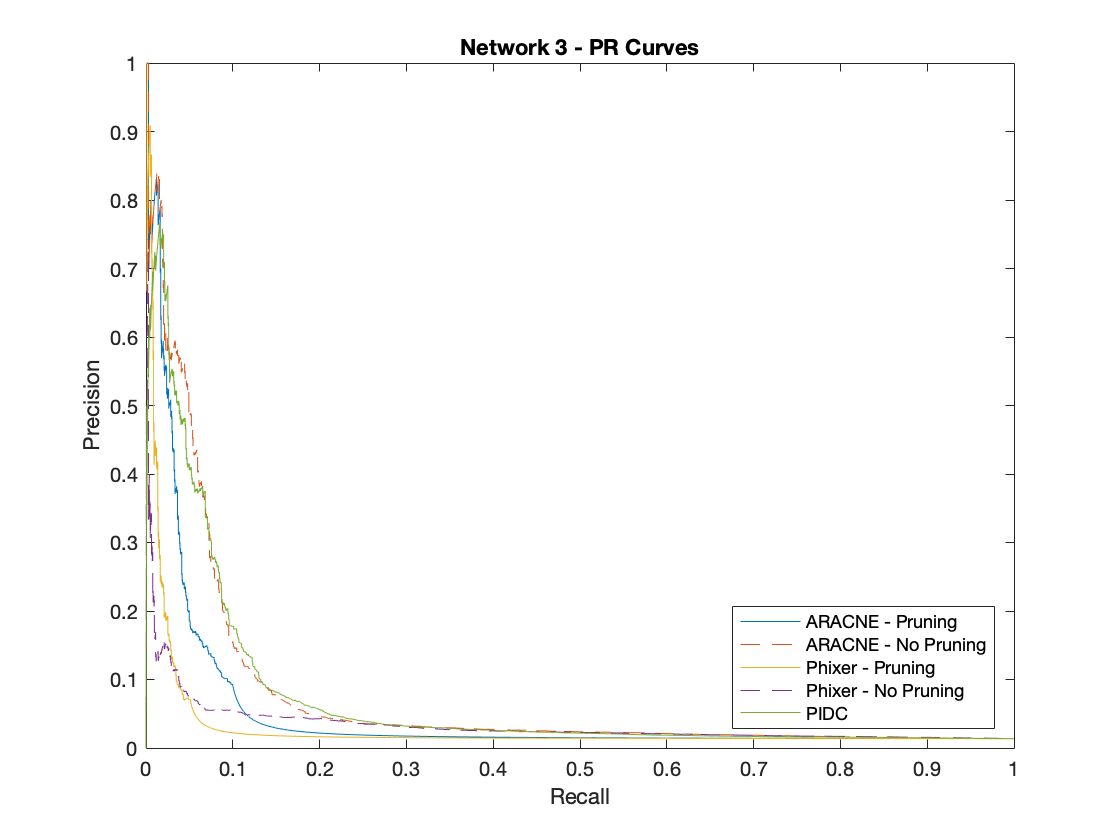}
      \caption{Network 3 PR curves.
}
      \label{figurelabel}
   \end{figure}

              \begin{figure}[thpb]

      \centering

      \includegraphics[scale=.23]{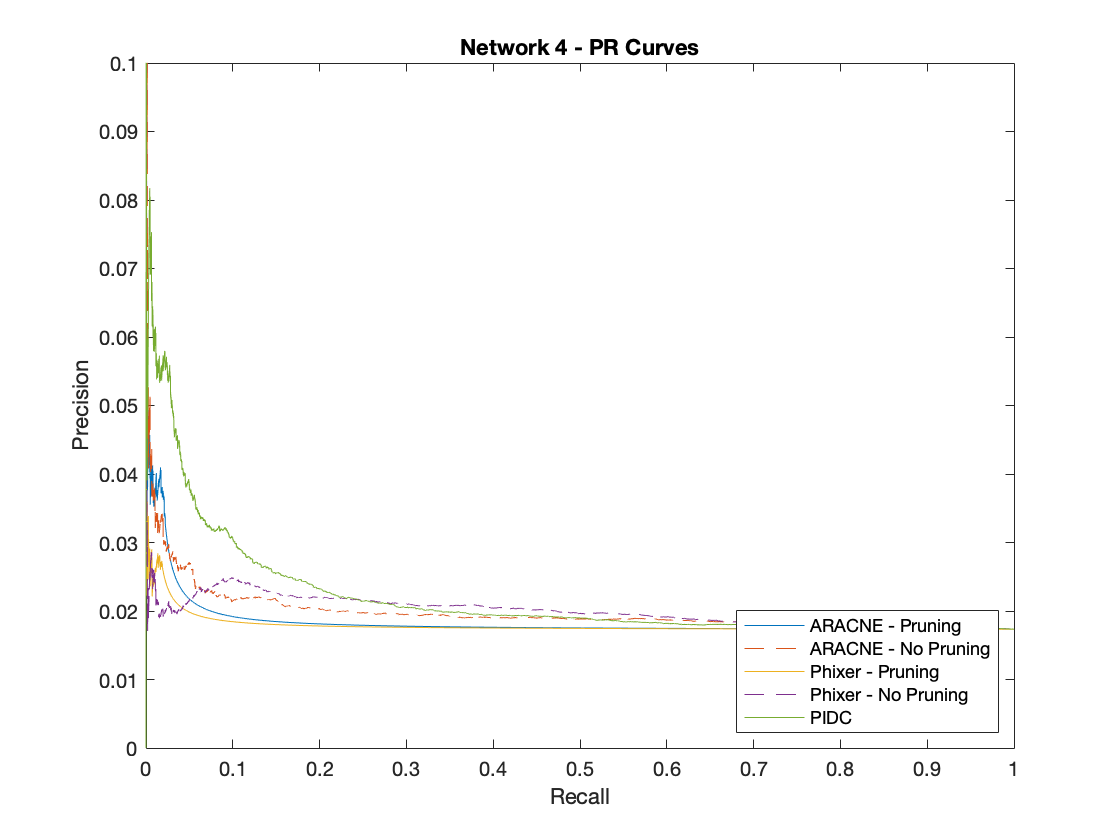}
      \caption{Network 4 PR curves.
}
      \label{figurelabel}
   \end{figure}

    The ROC analyses show that in each test case, the unpruned versions of ARACNE and Phixer achieved a higher recall than the pruned versions at nearly every possible false positive rate. In addition, the unpruned versions are able to compete with PIDC in terms of ROC curve accuracy, unlike the original pruned versions of the algorithms.
    
Figures 9, 10, and 11 show the PR curves for each network. Please note that the scale of the Y axis in Figure 11 is from 0 to 0.1, rather than from 0 to 1 as in the other figures. This is to make the differences between lines more visible, since Network 4 was very difficult and each algorithm achieved relatively low precision on it.
   
   While the results of the ROC analyses clearly favored the unpruned strategy, interpreting the results of the PR analyses requires more nuance. 
   
   Figure 9 illustrates an interesting trend in which the pruned strategy, for both ARACNE and Phixer, achieves a higher precision rate for low levels of recall, while the unpruned strategy achieves a higher precision rate for intermediate and high levels of recall. This general trend also appears in Figures 10 and 11, although it is more difficult to see visually due to the smaller differences in precision between the algorithms. 
   
   So, the pruned strategy may indeed be useful for certain situations in which the goal is to generate a simplified network prediction composed of a small subset of high-confidence interactions. However, if the goal is precision at an intermediate or high recall level, then the unpruned algorithm outperforms the pruned algorithm.

\subsection{Performance Summary - Unpruned vs. Pruned}

Table I shows the AUROC scores of the ARACNE and Phixer algorithms for all three gold standard networks, and the percentage change in AUROC from the unpruned strategy to the pruned strategy. Please note that in all six test cases, the unpruned strategy achieved a higher AUROC score than the pruned strategy.

\begin{table}[htbp]
\caption{ROC Curve Summary}
\begin{center}
\begin{tabular}{|c|c|c|c|c|}
\hline
\textbf{Algorithm}&\textbf{Dataset}&\multicolumn{3}{|c|}{\textbf{AUROC}} \\
\cline{1-5} 
& & \textbf{\textit{Unpruned}}& \textbf{\textit{Pruned}}& \textbf{\textit{\% Change}} \\
\hline
ARACNE& Network 1&0.7299 &0.6547&-10.30\%  \\
\hline
ARACNE& Network 3&0.6465 &0.5434&-15.94\%  \\
\hline
ARACNE& Network 4&0.5286 &0.5053&-4.40\%  \\
\hline
Phixer& Network 1&0.7848 &0.6157&-21.54\%  \\
\hline
Phixer& Network 3&0.6467 &0.5209&-19.45\%  \\
\hline
Phixer& Network 4&0.5401 &0.5033&-6.81\%  \\
\hline
\end{tabular}
\label{tab1}
\end{center}
\end{table}

\begin{table}[htbp]
\caption{PR Curve Summary}
\begin{center}
\begin{tabular}{|c|c|c|c|c|}
\hline
\textbf{Algorithm}&\textbf{Dataset}&\multicolumn{3}{|c|}{\textbf{AUPR}} \\
\cline{1-5} 
& & \textbf{\textit{Unpruned}}& \textbf{\textit{Pruned}}& \textbf{\textit{\% Change}} \\
\hline
ARACNE& Network 1&0.1874&0.2266 &+20.91\%  \\
\hline
ARACNE& Network 3&0.0745&0.0488 &-34.49\%  \\
\hline
ARACNE& Network 4&0.0199 &0.0185 &-7.03\%  \\
\hline
Phixer& Network 1&0.1417 &0.0984 &-30.55\%  \\
\hline
Phixer& Network 3&0.0337 &0.0289 &-14.24\%  \\
\hline
Phixer& Network 4&0.02 &0.0179 &-10.50\%  \\
\hline
\end{tabular}
\label{tab1}
\end{center}
\end{table}
Table II shows the AUPR scores of the ARACNE and Phixer algorithms for all three gold standard networks, and the percentage change in AUPR from the unpruned strategy to the pruned strategy. Please note that the unpruned strategy achieved a higher AUPR score than the pruned strategy in five out of six test cases. 

So, although pruning may be be useful in some contexts, as discussed in the previous section, the unpruned strategy appears to be the superior choice when the goal is overall accuracy, as measured by either the AUROC score or AUPR score.

\subsection{Comparison to PIDC}
To put our results into context, we tested PIDC [8] on the DREAM5 gold standard datasets and compared its accuracy to that of the unpruned ARACNE and Phixer algorithms. Table III shows the AUROC scores for all three algorithms on all three gold standard networks, while Table IV shows the AUPR scores.

\begin{table}[htbp]
\caption{AUROC Scores}
\begin{center}
\begin{tabular}{|c|c|c|c|c|}
\hline
\textbf{Dataset}&\textbf{ARACNE*}&{\textbf{Phixer*}}&{\textbf{PIDC}}&{\textbf{Winner}} \\
\hline
Network 1&0.7299 &0.7848 &0.7753 &Phixer  \\
\hline
Network 3&0.6465 &0.6467 &0.6313 &Phixer \\
\hline
Network 4&0.5286 &0.5401 &0.5314 &Phixer  \\
\hline
\multicolumn{5}{l}{*Unpruned strategy}
\end{tabular}
\label{tab1}
\end{center}
\end{table}

\begin{table}[htbp]
\caption{AUPR Scores}
\begin{center}
\begin{tabular}{|c|c|c|c|c|}
\hline
\textbf{Dataset}&\textbf{ARACNE*}&{\textbf{Phixer*}}&{\textbf{PIDC}}&{\textbf{Winner}} \\
\hline
Network 1&0.1874 &0.1417 &0.2255 &PIDC  \\
\hline
Network 3&0.0745 &0.0337 &0.0711 &ARACNE \\
\hline
Network 4&0.0199 &0.02 &0.0219 &PIDC  \\
\hline
\multicolumn{5}{l}{*Unpruned strategy}
\end{tabular}
\label{tab1}
\end{center}
\end{table}

From these results, we cannot conclusively say which is the best method of the three. However, it appears that once their pruning steps are removed, ARACNE and Phixer are at least competitive with PIDC, one of the most accurate algorithms in the field.

During the testing process, it also came to our attention that the unpruned ARACNE and Phixer algorithms outperform PIDC in terms of runtime. Runtimes for the three algorithms are shown in Table V and Figure 12. Measurements were taken on an OptiPlex 7060 Dell, with an Intel Core i7-8700 CPU and 7.6 GB RAM, running 64-bit Ubuntu 18.04.1. 

\begin{table}[htbp]
\caption{Runtime (minutes:seconds)}
\begin{center}
\begin{tabular}{|c|c|c|c|}
\hline
\textbf{Dataset}&\textbf{ARACNE*}&{\textbf{Phixer*}}&{\textbf{PIDC}} \\
\hline
Network 1&1:06 &14:11 &16:22  \\
\hline
Network 3&1:29 &107:31 &488:00 \\
\hline
Network 4&1:05 &114:09 &3398:44  \\
\hline
\multicolumn{4}{l}{*Unpruned strategy}
\end{tabular}
\label{tab1}
\end{center}
\end{table}

      \begin{figure}[thpb]
      \centering

      \includegraphics[scale=.65]{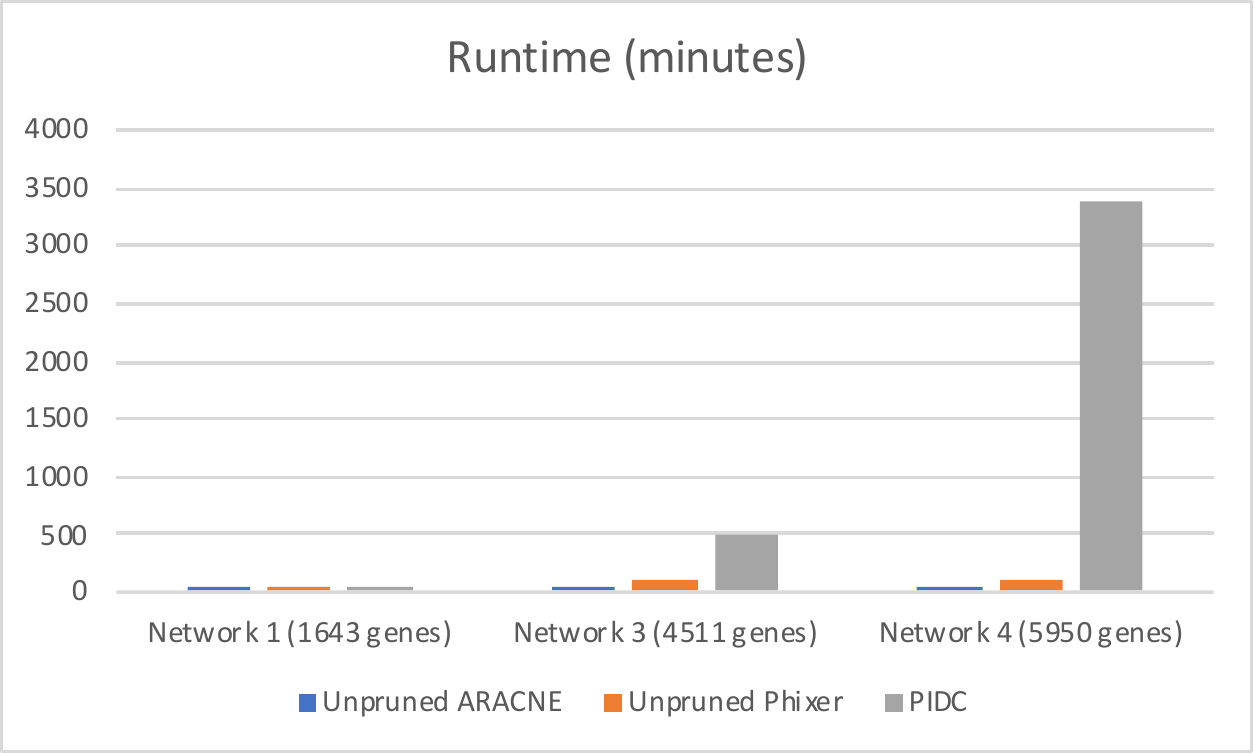}
      \caption{Runtime comparison.
}
      \label{figurelabel}
   \end{figure}  
   
In each case, ARACNE is the fastest and PIDC is the slowest, with Phixer in the middle. As the number of genes in the dataset increases, the difference becomes more evident. For Network 4 (the largest dataset, containing expression data for 5950 genes) ARACNE ran in about one minute, while Phixer ran in under two hours, and PIDC took more than two days. 

The difference in runtime scalability is likely due to the fact that ARACNE and Phixer perform edge weight calculations on pairs of nodes, while PIDC performs edge weight calculations on triplets of nodes. To give a simplistic example, consider a dataset with 100 genes. Since self-edges are not being considered, there are 100*99=9900 possible pairs of genes for ARACNE and Phixer to check, and 100*99*98=970200 possible triplets for PIDC to check. 

So, we believe that unpruned versions of ARACNE and Phixer can be considered viable alternatives to PIDC, since they deliver comparable levels of accuracy at faster runtimes, and scale better to large datasets.

\section{Discussion}

In this paper, we have shown that the pruning strategies used by ARACNE and Phixer reduced AUROC score in six out of six test cases and reduced AUPR score in five out of six test cases. While more research is needed to further confirm this, we tentatively conclude that these pruning methods have a negative overall effect on accuracy. Furthermore, we have shown that simply removing the pruning step allowed ARACNE and Phixer to attain accuracy levels similar to those of PIDC, a truly state-of-the-art inference algorithm.

However, we certainly do not mean to suggest that the pruning strategies are useless. Rather, it appears that they may be advantageous in some situations (such as when the goal is precision at a low recall level) and disadvantageous in others (such as when the goal is precision at an intermediate or high recall level). So, it is important for researchers to understand both the advantages and disadvantages of pruning in order to choose the appropriate inference strategy for their experimental context.

From a software development perspective, we suggest that inference software that includes a pruning step should also include an easy way to opt out of it, if the user so chooses. ARACNE has already done this, with the ``--nodpi'' option, but for Phixer we had to manually edit the code to remove the pruning step.

Next steps for this research will include analysis of other algorithms that include pruning strategies, such as CLR [9] and MRNET [10].

Further work will include an investigation into \textit{why} pruning leads to a drop in accuracy. We suspect that it is related to a loss of detection of redundant edges within feed forward loops, but this has yet to be confirmed. Once we can identify the root of the problem both empirically and theoretically, we will attempt to develop a new pruning strategy that achieves the goals of simplicity and false positive minimization without sacrificing accuracy.

%
%
%
%
%
%

\addtolength{\textheight}{-12cm}   






\begin{thebibliography}{99}

\bibitem{c1} Zhu X, Gerstein M, Snyder M (2007) Getting connected: analysis and principles of biological networks. Genes \& Development. 21:1010-1024

\bibitem{c2} Huynh-Thu VA, Sanguinetti G (2018) Gene regulatory network inference: an introductory survey. arXiv:1801.04087 [q-bio.QM]

\bibitem{c3} Madhamshettiwar et al. (2012) Gene regulatory network inference: evaluation and application to ovarian cancer allows the prioritization of drug targets. Genome Medicine. 4(5):41

\bibitem{c4} DREAM5 (2010) DREAM5 Challenge Website. Available online at: http://dreamchallenges.org/project/dream-5-network-inference-challenge/ 

\bibitem{c5} Margolin AA, Nemenman I, Basso K, Wiggins C, Stolovitzky G, Dalla Favera R, Califano A (2006) ARACNE: an algorithm for the reconstruction of gene regulatory networks in a mammalian cellular context. BMC Bioinformatics 7(1):S7. Available online at: https://github.com/califano-lab/ARACNe-AP/

\bibitem{c6} Singh et al. (2016) Inferring Genome-Wide Interaction Networks Using the Phi-Mixing Coefficient, and Applications to Lung Cancer. arXiv:1208.4006 [q-bio.GN]. Available online at: https://github.com/nitinksingh/phixer/

\bibitem{c7} Schrynemackers, Kuffner, Geurts (2013) On protocols and measures for the validation of supervised methods for the inference of biological networks. Frontiers in Genetics. 4:262


\bibitem{c8} Chan TE, Stumpf MPH, Babtie AC (2017) Gene Regulatory Network Inference from Single-Cell Data Using Multivariate Information Measures. Cell Systems. 5(3):251-267. Available online at: https://github.com/Tchanders/NetworkInference.jl


\bibitem{c9} Faith et al. (2007) Large-scale mapping and validation of Escherichia coli transcriptional regulation from a compendium of expression profiles. PLoS Biology. 5(1):e8

\bibitem{c10} Meyer PE, Kontos K, Lafitte F, Bontempi G (2007) Information-theoretic inference of large transcriptional regulatory networks. EURASIP Journal on Bioinformatics and Systems Biology. 2007(1):79,879

\bibitem{c11}Shen-Orr SS, Milo R, Mangan S, \& Alon U (2002) Network motifs in the transcriptional regulation network of Escherichia coli. Nature Genetics. 31(1), 64–68

\bibitem{c12}Milo R (2002) Network Motifs: Simple Building Blocks of Complex Networks. Science. 298(5594), 824–827

\bibitem{c13} Dekel E, Mangan S, \& Alon U (2005) Environmental selection of the feed-forward loop circuit in gene-regulation networks. Physical Biology. 2(2), 81–88














\end{thebibliography}
\end{document}